\documentstyle[12pt,a4,epsfig]{article}
\begin{document}
\newcommand{\be}{\begin{equation}}
\newcommand{\ee}{\end{equation}}        
\newcommand{\w}{wavelet}
\newcommand{\an}{analysis}
\newcommand{\co}{coefficient}
%\begin{titlepage}

\begin{center}
%\vfill

{\bf LONG-RANGE PARTICLE CORRELATIONS AND WAVELETS}

\vspace{2mm}

I.M. Dremin 
                    
\vspace{2mm}

{\it P.N.Lebedev Physical Institute, 117924 Moscow, Russia}

\vspace{2mm}
 
{\bf Contents}

\end{center}

\noindent 1. Introduction\\
2. About some long-range correlations\\
3. Earlier event-by-event studies\\
4. Wavelets: basic notions\\
5. The wavelet analysis of high-energy nucleus interactions\\
6. Conclusions\\
References\\

\begin{abstract}
The problem of long-range correlations of particles produced in high-energy
collisions is discussed. Long-range correlations involve large groups of
particles. Among them are, e.g., those correlations which lead to 
ring-like and elliptic flow shapes of individual high-multiplicity events in
the polar+azimuthal angles plane. The \w\ method of \an\ which allows to
disentangle various patterns has been proposed and applied to some central
lead-lead collisions at energy 158 GeV per nucleon. Previous attempts to find
out the ring-like correlations and recent results on \w\ \an\ of high-energy
nuclei interactions are reviewed.
\end{abstract}

\section{Introduction}

Recently, in Brookhaven National Laboratory (USA) started operating the
collider RHIC, where the colliding beams of the nuclei of heavy elements
interact with each other at the total energy up to 200 GeV per nucleon in the
center of mass system. There are some first interesting (even though still
preliminary) results, in particular, about the multiplicities of particles
produced. It happens, that, if recalculated  per a nucleon of the colliding
nuclei, it exceeds the multiplicity of nucleon-nucleon collisions at the same
energy. The events with the multiplicity of produced charged particles up
to 10000 become soon available. With the advent of the collider LHC in CERN
(Switzerland), when the energy of colliding nuclei will increase up to 1.8 TeV
per nucleon, the events with up to 20000 charged particles 
produced will be available. The problem of presentation and analysis of these
high multiplicity events becomes rather non-trivial. Each particle can be
represented by a dot in the 3-dimensional phase space. Therefore, the
distribution of these dots can be found for a single event. Different patterns
formed by these dots in the phase space would correspond to different
correlations and, therefore, to different dynamics. The problem of finding and
deciphering these patterns becomes crucial for understanding the underlying
dynamics. Really we are entering the new stage of studies of multiparticle 
production processes, where, apart from the traditional methods of selection
of definite events with the use of triggers or displaying the plots of the
simplest inclusive distributions, the exclusive characteristics of individual
events with high multiplicity become important in event-by-event \an\ . To
classify these patterns, one should develop the adequate methods of their
recognition and \an\ . The \w\ \an\ is just such a suitable method. Nowadays,
the very first (discussed in this paper) attempts to apply this method are
known, and it will, surely, find more wide applications.

Various correlations of particles produced in high-energy collisions are known.
Especially well studied are the two-particle correlations due to the decay
of two-particle resonances and due to the Bose-Einstein effect for identical
particles. The common correlation function technique, well known, for example,
in the statistical physics, is well suited for the \an\
of these two-particle correlations. It is more difficult to apply it if several
particles are involved because the correlation functions for three and more
particles depend on many variables and, therefore, are difficult to handle.
Nevertheless, correlations leading to clusters (or mini-jets, non-reducible to
resonances) and to jets (the correlated systems of many particles with a common
angular characteristics of their emission direction) have been studied as well,
however mostly in
$e^+e^-$-collisions where jet and subjet structures are clearly visible.
Otherwise one has to apply the averaging procedure to get knowledge of 
dynamical correlations from studies of the moments of various distributions and
their behavior in different regions of the phase space (for recent reviews, see
\cite{dwdk, koch, dgar}). It has lead to understanding of such global features
as the intermittent dynamics (first proposed in \cite{bpes}) and the fractal
structure of phase space distributions (first proposed in \cite{drev})
in these processes. It has been shown \cite{owoj, ddre, mpes} that, even at the
level of the lowest perturbative approximations, the quantum
chromodynamics (QCD) explains, at least
qualitatively, this structure as resulting from the intermittent emission
of subsequent jets with diminishing energies. Some completely new and unexpected
features such as oscillating cumulant correlators of the multiplicity
distributions were predicted in QCD \cite{dr93, dnec} and confirmed by
experiment (see, e.g., the review papers \cite{koch, dgar}). It demonstrated
the existence, besides the attractive correlations, also of repulsive
correlations between groups with different number of particles.

However, the genuine maltiparticle correlations originating from some collective
effects may still be hidden in individual events. It is not clear if they can
be disentangled with such averaging methods, especially if their probability
is not high enough. These local effects could be seen in some (probably, rare)
events as special patterns formed by produced particles within the available
phase space. They should be separated from statistical fluctuations leading to
similar patterns. There exists a method of the \w\ \an\ which allows to
recognize patterns due to correlations at different scales and damp down
the statistical noise even if it is very large in an initial sample
\cite{daub, meye, asta, dine}. The ability to reveal the local properties
of a process is the main advantage of the \w\ \an\ compared to other methods,
in particular, over the Fourier \an\ (for more detailed discussion see
\cite{dine}). In its turn, this article provides the more detailed review
of one of the problems briefly described in Ref. \cite{dine}.

\section{About some long-range correlations}

Before delving into applications of this method, let us consider examples of 
possible correlations which one would seek for. The well known example of
long-range correlations is provided, e.g., by the forward-backward 
correlations. However, they are of a global nature while we are more interested
here in such correlations which lead to special long-correlated patterns in
the available phase space. To be more definite, let us
concentrate on search for the so-called ring-like correlations which remind the
Cherenkov rings while others will be mentioned just by passing. Everybody
knows about the
Cherenkov radiation of photons by a bunch of electrons (or of any charged
particles, in general) traversing a medium whose refractivity index exceeds 1.
These photons form a ring in the plane perpendicular to electrons motion, i.e.
they are emitted at a definite polar angle. As a hadronic analogue, one may
treat an impinging nucleus as a bunch of confined quarks (color charges) each
of which can emit gluons when traversing a target nucleus. Let me remind that
Cherenkov photons are actually the collective effect of emission by the
medium which electrons traverse. In the same spirit, the Cherenkov gluons
would result from the collective emission by colliding nuclei. That is why
the ring-like structure, if confirmed, would indicate on the collective
long-range correlations in this system.

A long time ago, I speculated \cite{dr} about possible Cherenkov gluons relying
on experimental observation of the positive real part of the elastic forward
scattering amplitude of all hadronic processes at high energies. This is a
necessary condition for such process because in the commonly used formula for
the refractivity index its excess over 1 is proportional to this real part:
\be
\Delta n(\omega )=Re n(\omega )-1=\frac {2\pi N}{\omega ^2}Re A(\omega ,0^0)
\approx \frac {3\mu ^3}{8\pi \omega}\sigma (\omega )\rho (\omega ), \label{refr}
\ee
where $\omega $ is the energy of the emitted quantum, $N$ is the density of the
scattering centers which has been estimated for hadrons as
$N\approx 3\mu ^3/4\pi $ with
the pion mass $\mu $, and $A(\omega ,0^0)$ is the forward elastic scattering 
amplitude with $\rho (\omega )=Re A(\omega ,0^0)/Im A(\omega ,0^0)$ 
normalized by the optical theorem $Im A(\omega ,0^0)=\omega \sigma (\omega )/
4\pi $. Thus the necessary condition for the Cherenkov gluon radiation 
$\Delta n(\omega )>0$ is fulfilled if $\rho (\omega )>0$. For typical values
of $\sigma ,\rho $ and $\omega $ for hadronic processes one gets
$\Delta n(\omega )\ll 1$.

However, later \cite{dre2} I noticed that for such thin targets as hadrons or
nuclei the similar effect can appear due to small confinement length thus giving
us a new tool for its estimate. The bremsstrahlung, the Cherenkov gluons and the 
transition radiation in thin targets contribute to the same angular range
and thus become indistinguishable. At the same time, the difference between
bremsstrahlung and Cherenkov radiation can reveal itself in the increased
formation length of bremsstrahlung due to relativistic effects while for 
Cherenkov radiation the extension of a medium plays a decisive role. Therefore
bremsstrahlung will be increased and inclined to smaller polar angles. Thus the
background to Cherenkov radiation seemingly increases. However, as we shall see,
it tends to large polar angles in the center of mass system what favours its
observation. 

The general formula \cite{ftam} for the total energy $dW$ in the solid angle
$d\Omega $ due to emission of vector particles with energies in the interval
from $\omega $ to  $\omega +d\omega$ by a step-like
charge current $\vert {\bf j}_{\mu}^{\perp }\vert $ traversing a target of the
thickness $l$ and of the refractivity index $n$ with the velocity
$v=\beta c$ looks as
\be
\frac {dW}{d\Omega d\omega}=\frac {e^2\beta ^2}{\pi ^2c}\frac
{\sin ^2[\omega l(1-\beta n\cos \theta)]}{(1-\beta n\cos \theta)^2}\sin ^2\theta ,
             \label{ftam}
\ee 
where $e$ is the electric charge, $\theta $ is the polar angle of photons emission.

Such a step-like color current was used in Refs. \cite{dr, dre2} as a model
for deconfinement of quarks (color) of impinging hadrons (nuclei) only 
within the volume with the strong interaction radius $l\sim \mu ^{-1}$.
Outside this region, quarks are confined, and there are no color currents.
Thus the phenomenological parameter $l$ describes the range of confining
forces in usual static hadrons which just corresponds to the deconfinement
region during the collision and wherefrom this collective effect is observed.
The spin and mass properties of quarks and gluons are similar
to those for electrons and photons, correspondingly. Therefore,
to proceed to emission of high energy gluons by quarks deconfined in some
limited volume (step-like color current)  
from the formula (\ref{ftam}), one should replace there $\alpha $ by
$\alpha _SC_F$, where $\alpha _S$ is the QCD running coupling, $C_F=4/3$
and take into account the smallness of the emission angle in the laboratory
system $\theta _{l.s.}$, of the differences $\Delta n=n-1$ and $1-\beta $
as well as of the ratio $\mu /\omega $. Then one gets \cite{dr, dre2}
from (\ref{ftam}) the inclusive cross section for this gluon radiation
\be
\frac {\omega }{\sigma }\frac {d^3\sigma }{d^3k}=\frac {4\alpha _SC_F
\theta _{l.s.} ^2\sin [\omega l(\theta_{l.s.} ^2-2\Delta n)/4]}{\pi ^2
\omega ^2(\theta _{l.s.} ^2-2\Delta n)^2}.
\label{dre2}
\ee
In the case of the electromagnetic radiation in macroscopic targets, the 
condition $\omega l\Delta n/2\gg 1$ is fulfilled, and the righthand side
becomes the Dirac delta-function with an argument corresponding to the
condition for the usual Cherenkov rings of photons. In the hadronic case,
the opposite inequality
\be
\omega l\Delta n/2\sim 3\rho(\omega )/16\pi \ll 1   \label{ineq}
\ee
(for $\sigma \sim \mu ^{-2}, \;\;   l\sim \mu ^{-1}$) is valid. Thus the
position and the shape of the ring as defined by (\ref{dre2}) is determined 
by the confinement length $l$ but not so much by $\Delta n$.

Surely, the confinement of hadronic
partons leads to additional screening of low-energy gluon radiation which
becomes of the multipole nature because the initial current (a hadron or a 
nucleus) is colorless in distinction to a charged current of an electron bunch.
Thus this gluon radiation should be even of more collective nature and with
harder spectrum than
usually ascribed to the Cherenkov photon radiation. However it is hard to
account for this collective behavior in a more quantitative way.
The damping due to the imaginary part of the amplitude was
estimated \cite{dre2} as unimportant and given by the factor close to 1
\be
\exp [-\frac {3\mu ^2\sigma l}{8\pi }]\approx \exp [-\frac {\mu l}{8}]
\approx 0.9,         \label{damp}
\ee
which does not prevent the observation of this effect if any.

If several gluons are emitted and each of them generates a mini-jet centered
at a definite polar angle (or pseudorapidity $\eta =-\ln \tan \theta /2$
) without any condition imposed on
its azimuthal angle $\varphi $, the ring-like substructure will be
observed in the target diagram, i.e., in the plane perpendicular to the
collision axis. Therefore, the events possessing such
substructure with relatively short rapidity
correlation length and large azimuthal correlation length would be important
to look for. If the density of mini-jets within the ring is so high that they
overlap, then they form a (circular) ridge pattern (or a wall pattern
according to \cite{pesc}). If the number
of emitted gluons is not large, we will see several (or just one) jets (tower
structure \cite{pesc}) correlated in their polar, but not in the azimuthal
angle. The formula (\ref{dre2}) predicts quite large polar angle of emission of
gluons, and therefore quite large radius of
the ring in the target diagram that favors its observation 
\be
\theta _{l.s.}^{max}\geq \sqrt {2\pi /El},     \label{angl}
\ee
where $E$ is the collision energy. It gives for the deconfinement length of
the order of extension of nuclear forces $l\sim \mu ^{-1}$ the following
estimate of the polar angles for the rings in the center of mass system for
two identical particles
\be
\theta _{c.m.s.}\sim  70^0 \;\; (110^0),  \label{thet}
\ee
where the angle in the brackets corresponds to the emission in the backward
hemisphere by another colliding particle.
If the radiation length for colliding nuclei is proportional to $A^{1/3}$,
then the corresponding radiation angle should behave as $A^{-1/6}$.
It can be checked in collisions of nuclei with different atomic weights.
In a single event, either both or one of these rings can be formed depending
on the probability of such a process.

Surely, for the parton-emitter moving already at some angle to the collision
axis this ring will be transformed into ellipse.
Central collisions of nuclei would be preferred for observation of 
such effects because of a large number of participating partons though the
background due to ordinary processes increases as well. If the number
of correlated particles within the ring is large enough, it would result in
spikes in the
pseudorapidity distributions. However, the usual histogram method is not always
good to verify these spikes because it may split a single spike into two bins
thus diminishing its role. However, some hints to such structure can be found
from these histograms.

Namely, such a histogram of a cosmic ray nuclear 
interaction \cite{addk} initiated my approach to this problem. A huge spike on
the rapidity plot observed in this event was initially interpreted as resulting
from a single cluster. However, since the number of particles in the spike was
very high (56 charged particles contributed to it while the average number is
about 10) it was more carefully studied, and it was found that they
form a ridge in the target diagram (the polar+azimuthal angles plane).
There was also indication on another more diluted ring-like structure in
the backward hemisphere of the same event.
High spikes in the pseudorapidity distributions have been observed in some
other cosmic ray data \cite{alex, masl, arat, dtre} and in event-by-event
\an\ of accelerator data \cite{maru, adam}. Especially impressive is the 
famous event of NA22 Collaboration \cite{adam} of the pion-nucleon interaction
with a spike 60 times
exceeding the average density in the narrow rapidity window. The particles
inside the spike are rather uniformly distributed in azimuthal angles
(the ridge structure). However, the multiplicities in the analyzed central
nucleus-nucleus events are about 50 times higher than in accelerator data
on hadron-hadron collisions.
In this case the huge combinatorial background may dilute the strength of
spikes. Anyway all these events were just single representatives chosen by
eye from samples of other events.

It is usually argued that the \an\ of a
single event of the central nucleus-nucleus collision at high energies may be
statistically reliable due to a large number of particles produced.
To classify such events in a more quantitative way, one should have a precise
method of the local pattern recognition at different scales with a possibility
to use an inverse transform. It became possible with the advent of the
recently developed methods of the \w\ \an\ .
The \w\ \an\ is well suited for this purpose because it
clearly resolves the local properties of a pattern on the event-by-event basis.

Another example of collective long-range correlations is provided by the
so-called elliptic flow i.e. the azimuthal
asymmetry in individual events. It may be related to a collective classical
sling-effect \cite{dman} of the rotation of colliding nuclei after peripheral
collisions initiated by the pressure \cite{olli} at some impact parameter at
the time of collision. It can give some knowledge about the equation of state
of the hadronic matter by studies of the shapes of the created squeezed states.
Another origin of the effect could be due to some jetty structures because
emission of two jets with
high transverse momenta well balanced by energy-momentum conservation laws
would result in the azimuthal asymmetry of an individual event. These two
theoretical suppositions lead however to different event patterns and can be
distinguised in experiment. By passing, let
me say that the elliptic flow patterns ("cucumber") corresponding to the large
value of the second Fourier coefficient, and the "three leaves flower"
pattern with large third coefficient were also observed when scanning some
high multiplicity events. These results have not been published yet, and I do
not discuss all these effects here concentrating discussion on the ring-like
events only.

The event-by-event \an\ of patterns in experimental and Monte Carlo events
becomes especially important for very high-multiplicity collisions at RHIC and
LHC. The homogeneous 4$\pi $ acceptance of detectors (such as that of STAR
Collaboration at RHIC) will be crucial for it, not to provide false patterns.
I am
sure that various patterns will be observed and allow triggering on different
classes of events and classifying "anomalous" features. Let me stress, however,
here that the background due to the ordinary processes of parton emission and
rescattering is huge in nucleus-nucleus collisions. It is not at all clear if 
the described above collective effect will be noticeable and can be separated
from more common traditional patterns of radiation by individual quarks and
gluons. One may rely only on very specific features of this effect for its
unanimous registration and on first positive experience in this respect.
Therefore, the clear separation and recognition of these (low-probability?)
patterns as objective dynamical effects requires new identification methods
insensitive to the smooth background and statistical fluctuations (noise).
As one of them, I propose to use the \w\ \an\ and describe recent developments
in this direction briefly reviewing first some other proposals.

\section{Earlier event-by-event studies}

When the target diagrams of individual events are imaged visually, the human
eye has a tendency to observe different kinds of intricate patterns with dense
clusters (spikes) and rarefied voids in the available phase space. However,
the observed effects are often
dominated by statistical fluctuations and look quite subjective. The method of
factorial moments was proposed \cite{bpes} to remove the statistical
background in a global \an\ and it shows fractal properties even in
event-by-event approach (see \cite{dwdk}). The increase of the factorial
moments at small bins signals the presence of non-statistical fluctuations.
Thus this method may be used as a tool for the preliminary selection of
events with strong dynamical fluctuations. Nevertheless, it averages somehow
the information about event patterns. Moreover, the patterns in the target
diagrams as seen by eye hardly differ sometimes \cite{cddh} in the events
with different factorial moments behavior. Some more sensitive and selective
criteria should be used.

First detailed event-by-event \an\ 
\cite{dlln, agab} of large statistics data on hadron-hadron interactions
(unfortunately, however, for rather low multiplicity) was performed to look for
the dense groups of particles well separated (isolated) from other particles in an event.
The dense groups could imitate single dense jets or ring-like events. Some 
threshold values were imposed from below on the density of the groups and on
their rapidity distance from other particles. These groups were quite narrow.
The rapidity locations of their centers were determined. It has been found that
the centers of these groups prefer to be positioned at a definite
polar angle. The positions of the maxima of the centers distribution are quite
close to estimates according to the formula (\ref{thet}). This feature favors
the above interpretation in terms of Cherenkov gluons.

For nucleus-nucleus collisions, first systematic event-by-event \an\ was
attempted by NA49 Collaboration \cite{rola}. Unfortunately, it was limited only
by studies of the fluctuations in the particle transverse momenta and the 
relative production of kaons to pions. No evidence was found for unusual
fluctuations in the ratios of kaons to pions and in the fluctuations of the
transverse momenta even though the latter were much smaller than in 
nucleon-nucleon collisions that can be treated on a qualitative level as a
result of intra-nuclear
rescatterings. These conclusions do not tell us anything about patterns in
individual events. It is not surprising because this experiment has a limited
and inhomogeneous acceptance, and only a fraction of the secondary particles
is actually recorded. Moreover, this feature of the detector distorts some
event characteristics crucial for pattern recignition, e.g., such as the
particle density fluctuation. Therefore its results are not directly useful
for our purposes.
 
The full space coverage is yet ensured only in the traditional emulsion chamber 
experiments. Therefore it is worthwhile to attempt the event-by-event \an\
of their data. Even though the total number of events is relatively low and
the particle identification is limited, they are prominent for their high
angular resolution which is important for studies of the particle density 
fluctuations. Very interesting systematic event-by-event \an\ of 
high-multiplicity Pb--Ag/Br collisions at energy\footnote{Up to now, the
nuclear collisions were studied in the external beams only. Therefore
everywhere we imply the energy per nucleon of the impinging nucleus in the
rest system of the target nucleus. The impinging nuclei are shown first in
front of the defis sign, and the target nuclei afterwards. It is clear that
collisions of different nuclei at colliders will ask for other conditional
notations.} 158 GeV detected in EMU13
emulsion experiment was performed by the KLM Collaboration \cite{cddh}.
The results were confronted to three different Monte Carlo models:
Fritiof, Venus and the random-type model called SMC. It was noticed that
even in the one-dimensional distributions the probability to find a spike
and sizes of spikes are systematically larger in measured events than in
simulated ones. It was shown by plotting the distributions of spike
fluctuations. The combinatorial background dilutes the strength of the
observed signals. On the two-dimensional $\eta -\varphi $ phase space plot
the slightly stronger clustering (jettyness) of particles in the measured
events was also observed, especially for small size clusters. The size of the
cluster (cone) was defined as $R=\sqrt {\delta \eta ^2+\delta \varphi ^2}$ and
the number of particles in the cone was chosen greater than 4. The fraction
of particles confined in clusters is quite large. No rapidity and azimuthal
angle distributions of the cone centers were presented, unfortunately.
Therefore one can not decide if this effect is only due to short range 
correlations or some long range correlations like those discussed above are
important as well. The only general conclusion is that the phase space 
inhomogeneity (jettyness) is stronger in measured events of nucleus-nucleus
collisions than in any of Monte Carlo models based on conventional physics
of nucleus-nucleus collisions.

In the Ref. \cite{aaaa} the azimuthal substructure of particle distributions
in individual central high-energy heavy-ion collisions within dense and
diluted groups of particles along the rapidity axis was investigated. The
data of EMU01 Collaboration on O/S - Ag/Au collisions at 200 GeV  
were used. Some criteria appeared to be rather insensitive and did not show
any significant difference from the stochastic averages with
$\gamma $-conversion and HBT particle interference effect taken into account.
However, when the parameter
\be
S_2=\sum _i(\Delta \varphi _i/2\pi )^2,   \label{stwo}
\ee
where $\Delta \varphi $ is the azimuthal difference between two neighboring
particles in the group and sum is over all particles $i$ in the group, was used,
it revealed some jet-structure for the
dilute groups which was impossible to explain by known effects. It is an 
indication on the ridge-like structure (not a tower structure!) of analyzed
groups within the definite rapidity windows. This \an\ underlines the 
importance of the choice of criteria sensitive enough to features which may
be hidden in particular patterns. To my opinion, even the parameter $S_2$
when used in the "histogram-like" approach with a fixed scale length in
pseudorapidity $\Delta \eta $ as was done in Ref. \cite{aaaa} averages too
much the fluctuations in individual events. More local characteristics should
be used not to smear the distinct differences between different patterns.

The dense groups were studied in nucleus-nucleus collisions at lower energies.
Their rapidity distributions also showed \cite{ggsa} some peaks similar to 
those found in Refs. \cite{dlln, agab}. However, the fragmentation processes
are so strong here that it is not clear how the fragment products influence
this conclusion.

\section{Wavelets: basic notions}

All these results, valuable by themselves, do not still answer the question
about the existence of long-range correlations in individual events. To get it,
one should be able to perform the large-scale event-by-event \an\ . Such a
tool is provided by \w s. Let us briefly describe basic notions about \w s
(for more details, see \cite{asta, dine}).

Commonly used wavelets form a complete orthonormal
system of functions with a finite support by using dilations and translations.
 That is why by changing a scale (dilations) they can distinct the local 
characteristics of a signal at various scales, and by translations they 
cover the whole region in which it is studied. The orthogonality of \w s
insures that the information at a definite resolution level (scale) does not
interfere with other scale information. The \w\ \an\ is the study of any
function by expanding it in the \w\ series (or integrals). Due to the
completeness of the system, they also allow for the inverse transformation
(synthesis) to be done. It means that the original function or some parts
of it containing the investigated correlations may be restored without any
loss of the information. In the \an\ of nonstationary signals or inhomogeneous
images (like modern paintings with very sharp figure edges), the locality
property of \w s leads to their
substantial advantage over Fourier transform which provides us only with the
knowledge of global frequencies (scales) of the object under investigation
because the system of functions used (sine, cosine or complex exponents) is 
defined on the infinite interval.

The \w\ \an\ reveals the {\it local} properties of
any pattern in an individual event at {\it various} scales and, moreover, avoids
smooth polynomial trends and underlines the fluctuation patterns. By choosing
the strongest fluctuations, one hopes to get rid of statistical fluctuations and
observe those dynamical ones which exceed the statistical component. 

The traditional formula for the \w\ transform of a one-dimensional function
$f(x)$ is written as
\be
W(a,b)=a^{-1/2}\int f(x)\psi (\frac {x-b}{a})dx,   \label{wabf}
\ee
where $\psi $ denotes a \w\ with its argument shifted to $x=b$ (translation)
and scaled by $a$ (dilation). For continuous \w s, both $a$ and $b$ are
continuous variables. For discrete \w s, one usually chooses $a=2^j$ where
$j$ are integer numbers and replaces $b$ by $2^jk$. As we see, the \w\ \co s
for a one-variable function are the functions of two variables instead of one
in case of the Fourier transform. It allows now to define both the scale
(frequency) and its effective location.

The choice of the \w\ depends on the problem studied and is not unique. As an
example of continuous \w s, let us mention the so-called "Mexican hat" \w\
which is nothing else as the second derivative of the Gaussian function. The
discrete \w s are obtained as solutions of a definite functional equation and
cannot be represented in the analytical form. However, they are suitable for
computer calculations (for more detail see \cite{asta, dine}).

\section{The wavelet \an\ of high-energy nucleus interactions}

First attempts to use \w\ \an\ in multiparticle production go back to
P. Carruthers \cite{carr, lgca, gglc} who used \w s for diagonalisation of 
covariance matrices of some simplified cascade models. The proposals of
correlation studies in high multiplicity events with the help of \w s were 
promoted \cite{sboh, huan}, and used, in particular, for special correlations 
typical for the disoriented chiral condensate \cite{shth, nand}. The
\w\ transform of the pseudorapidity spectra of JACEE events was done in Ref. 
\cite{sboh}.

As was mentioned above, \w s are most effective in the \an\ of inhomogeneous
patterns. That is why I proposed to use them for deciphering the phase
space inhomogeneity (in particular, the target diagrams) of very high
multiplicity events.

At present, only five high-multiplicity events of the lead-lead collisions at
158 GeV were analyzed according to this method \cite{adk, dikk}. I
demonstrate here these central Pb-Pb events with the highest registered
multiplicities from 1034 to 1221 charged particles chosen from 150 processed
events and used for wavelet decomposition with the aim to study the patterns
inherent to them. The data were taken from the emulsion chamber (with the thin
lead target) experiment EMU15 at CERN by the group from Lebedev Physical Institute.

The target diagrams of secondary particles distributions for these events
\cite{dikk} are
shown in Fig. 1, where the radial distance from the center measures the polar
angle $\theta $, and the azimuthal angle $\varphi $ is counted around the center.
I show them here to demonstrate that even if one really notices some
inhomogeneities in these diagrams, it is not easy to claim which one is of the
dynamical origin and, moreover, the whole pattern is strongly influenced by
the trivial high energy effect of higher density at small polar angles.
One can sum over the azimuthal angle and plot the corresponding pseudorapidity
($\eta =-\log \tan \theta /2$) distributions shown in Fig. 2. The pronounced 
peaks ($\eta $-spikes) strongly exceeding expected 
statistical fluctuations are seen in individual events. This inhomogeneity in 
pseudorapidity can arise either due to a very strong jet i.e. a large group 
(tower) of particles close both in polar and azimuthal angles or due to a
ring-like (ridge) structure when several jets with smaller number of particles
in each of them have similar polar angle but differ in their azimuthal angles.
However, it is still not easy to observe them directly in the target diagram even
if the pseudorapidity is plotted along its radius instead of the polar angle
or the scaled variable \cite{bgad}
\be
\bar {\eta } (\eta )=\int _{\eta _{min}}^{\eta }\rho(\eta ')d\eta '/
\int _{\eta _{min}}^{\eta _{max}}\rho(\eta ')d\eta ',  \label{bgad}
\ee
which gives rise to a flat distribution $\rho (\bar {\eta })$, is used.

In Ref. \cite{adk} \w s were first used to analyze two-dimensional 
patterns of fluctuations in the phase space of a single event, shown in
Fig. 1 at number 19. There were presented \cite{adk} the results of
the {\it one-dimensional} analysis of separate sectors of this two-dimensional
plot. To proceed in this way, the whole azimuthal region was divided into 24
sectors of the $\pi /12$ extension each (thus preserving, unfortunately, the
histogram drawback in this coordinate). The pseudorapidity distributions in
each of them were separately analyzed after integrating over azimuthal angles
within a given sector. Neighboring sectors were connected afterwards.

Both jet and ring-like structures have been found
from the values of squared wavelet coefficients as seen from Fig. 3 taken from 
Ref. \cite{adk}. Four of 24 sectors are demonstrated there. The \w\ \co s
were calculated using the continuous "Mexican hat" \w\ . The darker regions
correspond to larger fluctuations, i.e. to larger values of \w\ \co s. At the
very top in each box, i.e. at small scales $a$, the wavelet analysis
reveals individual particles as they are placed on the pseudorapidity axis
in a given sector seen as short dark lines. At larger scales $a$ corresponding
to the shift down in any box of the Figure,
the correlations of different shapes leading to clusters or jets of particles
are resolved. Finally, at ever larger scale one notices the long-range
structure (indicated by the arrows in Fig. 3)
which penetrates from one azimuthal sector to the neighboring one at nearby
values of the polar angle (pseudorapidity), thus forming an elliptic ridge
around the center of the target diagram. This structure approximately
corresponds to the peak in the pseudorapidity distribution (for more detail,
see Ref. \cite{adk}). Let us note, that it is not easy to notice in the target
diagram of this event, shown in Fig. 1, any increase of
density at the ring just by eye because of the specific properties of the
$\theta - \varphi $ plot where the density of particles decreases fast toward
the external region of large polar angles due to the relatistic effects.
 Also, the tree-like patterns in
this plot of the \w\ \co s would probably correspond to the fractal structure
of the phase space which was discovered by the factorial moments method. 
 
To reveal these patterns in more detail one should perform the
{\it two-dimensional}  local analysis. It is strongly desirable to get rid of
such drawback of the histogram method as fixed positions of bins that sometimes
gives rise to splitting of a jet into pieces contained in two or more bins as,
e.g., it could happen in the above \an\ of the event 19 \cite{adk} when 24
azimuthal sectors
were chosen. Moreover, the number of charged particles in each sector was
about 50, that is already at the limits of the accuracy for efficient \w\ \an\
to be done. The two-dimensional wavelet transform of particle densities
directly in the two-dimensional plot does not have such a deficiency.
Wavelets choose automatically the size and shapes of 
bins (so-called Heisenberg windows; see, e.g., Ref. \cite{daub}) depending on 
particle densities at a given position. The multiresolution
analysis at different scales and in different regions is performed.

In principle, the wavelet coefficients $W_{j_1,k_1,j_2,k_2}$ of the
two-dimensional function $f(\theta ,\varphi )$ are found from the formula
\begin{equation}
W_{j_1,k_1,j_2,k_2}=\int f(\theta ,\varphi )\psi (2^{-j_1}\theta -k_1;2^{-j_2}
\varphi -k_2)d\theta d\varphi .   \label{wav}
\end{equation}
Here $\theta _i, \varphi _i$ are the polar and azimuthal angles of particles produced,
$f(\theta ,\varphi )=\sum _i\delta (\theta -\theta _i)\delta (\varphi -\varphi _i)$
with a sum over all particles $i$ in a given event, $(k_1,k_2)$ denote the 
locations and $(j_1,j_2)$ the scales analyzed. The function $\psi $ is the 
analyzing wavelet. The higher the density fluctuations of particles in a given 
region, the larger are the corresponding wavelet coefficients.

In practice, the discrete wavelets obtained from the tensor product of
two multiresolution analyses of standard one-dimensional Daubechies 8-tap
wavelets were used. Then the corresponding $ss, sd$ and $dd$ coefficients in
the two-dimensional matrix were calculated (see \cite{daub}). The common scale
$j_1=j_2=j$ was used. It simplifies the calculations and the presentation of
obtained results because of the smaller number of variables but may be not
completely satisfactory from physical point of view and, probably, should be
abandoned later in a more sofisticated approach. The similar basis has been
used in Ref. \cite{goiv}.

As stressed above, the ring-like structure should be a collective effect
involving many particles and large scales. Therefore,
to get rid of the low-scale background due to individual particles and analyze
their clusterization properties, the scales $j>5$ have been chosen where both
single jets and those clustered in ring-like structures can be revealed as
seen from Fig. 3. Therefore all coefficients 
with $j<6$ are put equal to 0. The wavelet coefficients for any $j$ 
from the interval $6\leq j\leq 10$ are now presented as functions of 
polar and azimuthal angles in a form of the two-dimensional landscape-like 
surface over this plane i.e. over the target diagram. Their inverse wavelet 
transform allows to get modified target diagrams of analyzed events with
large-scale structure left only. Higher fluctuations of particle density inside
large-scale formations and, consequently, larger wavelet coefficients
correspond to darker regions in this modified target
diagram shown in Fig. 4. Here we demonstrate two events (numbered 3 and 6) from
those five shown in Figs. 1 and 2. They clearly display both jet and ring-like 
structures which are different in different events. To discard
the methodical cut-off at $\eta \approx 1.6-1.8$ the region of $\eta >1.8$
was only considered\footnote{The more detailed discussion about this
methodical effect can be found in Ref. \cite{dikk}. It shows that \w s may be
used for the \an\ of methodical problems in detectors as well. This topic is
out of the scope of the present paper.}.

Even though the statistics is very low, it was attempted to plot the 
pseudorapidity distribution of the maxima of \w\ coefficients with the
hope to see if it reveals the peculiarities observed in high statistics but
low multiplicity hadron-hadron experiments \cite{dlln, agab}. In Fig. 5, the
number of highest maxima of wavelet coefficients exceeding the threshold value
$W_{j,k}>2\cdot 10^{-3}$ is plotted as a function of their pseudorapidity for
all five events considered. It is quite peculiar that positions of the maxima
are discrete. They are positioned quite symmetrically about the value 
$\eta \approx 2.9$ corresponding to $90^0$ in the center of mass system as it
should be for two Pb nuclei colliding. Difference of heights is within the
error bars.
More interesting, they do not fill in this central region but are rather
separated. Qualitatively, it coincides with findings in Refs. \cite{dlln, agab},
where the separated peaks in the pseudorapidity distribution of the centers of
dense groups were also noticed approximately at the same positions.

For comparison, there were generated 100 central Pb-Pb interactions with energy 
158 GeV according to Fritiof model and the same number of events
according to the random model describing the inclusive rapidity distribution
shape. The fluctuations in these simulated high multiplicity events are much
smaller than in experimental ones and do not show any ring-like structure.

Further use of \w s by imposing different scales for polar and azimuthal angles
and study of the 3-dimensional phase space are desirable. For the latter, one
needs the data on the momenta of the created particles, which may be obtained
only in experiments with the magnetic field. Though the emulsion chamber of
the experiment EMU15 was installed in the magnetic field $H=1.8$T, there
were no measurements of the trajectory curvature done until now. Thus the data
on the three-dimensional phase space are not yet available. I would like to
stress once more that good homogeneous acceptance of the detectors with 4$\pi $
geometry is crucial for the proper \w\ \an\ of any event to get firm conclusions
about the dynamical effects. Otherwise the results will show an admixture of
physical and methodical effects which can not be disentangled in the
event-by-event \an\ . Actually, this remark is a byproduct of my own 
experience in attempts to find out the elliptic flow effect in NA49 data on
lead-lead collisions at 158 GeV. They failed because I noticed the azimuthal
asymmetry which persisted at the same position, suspected it to be due to
the inhomogeneous azimuthal acceptance of the detector, and after request was
insured by the authors that my guess is correct. Thus the methodical effect was
so strong that it prevented any conclusions about the dynamical elliptic flow
but, probably, it will not prevent from the \an\ of some strucnures in polar
angles to be done. Even though such applications of the \w\ \an\ are interesting
by themselves and their effectiveness was demonstrated in this case,
they are not of the main physics interest and of our concern here.

\section{Conclusion}

Thus I conclude that even on the qualitative level there is the noticeable 
difference between experimental and simulated events with larger and somewhat 
ordered fluctuations in the former ones. First attempts to apply the method of
the \w\ \an\ described here have shown some peculiar patterns of individual
measured events not resolved in the Monte Carlo simulated samples. It means
that we still have something to learn about their dynamical features. 
The ring-like patterns were used here only as an example of possible
correlations leading to some special particle grouping in the phase space.
More detailed \w\ \an\ at various resolution levels and also in the three
dimensional phase space may reveal some yet undiscovered structures. The
described above efforts show that there are no unsolvable problems for doing it.

My aim here is to show the
applicability and power of a new method of the two-dimensional \w\ \an\ , the
qualitative features and differences leaving aside quantitative characteristics
till higher statistics of fully registered high multiplicity AA-events becomes
available. In particular, the special automatic complex for emulsion processing
with high space (angular) resolution (www.lebedev.ru/structure/pavicom/index.htm)
in Lebedev Physical Institute is coming into operation, and it can help
enlarge the statistics of analyzed central Pb-Pb collisions at 158 GeV
quite soon using at full strength the good acceptance and precision of emulsion
detectors. Also the data of STAR collaboration in RHIC with a full phase space
coverage and large number of events should soon be available at ever
higher energies. Wavelets provide a powerful tool for event-by-event analysis
of fluctuation patterns in such collisions.

\newpage

\begin{center}
\vspace*{-3.2cm}
\epsfig{file=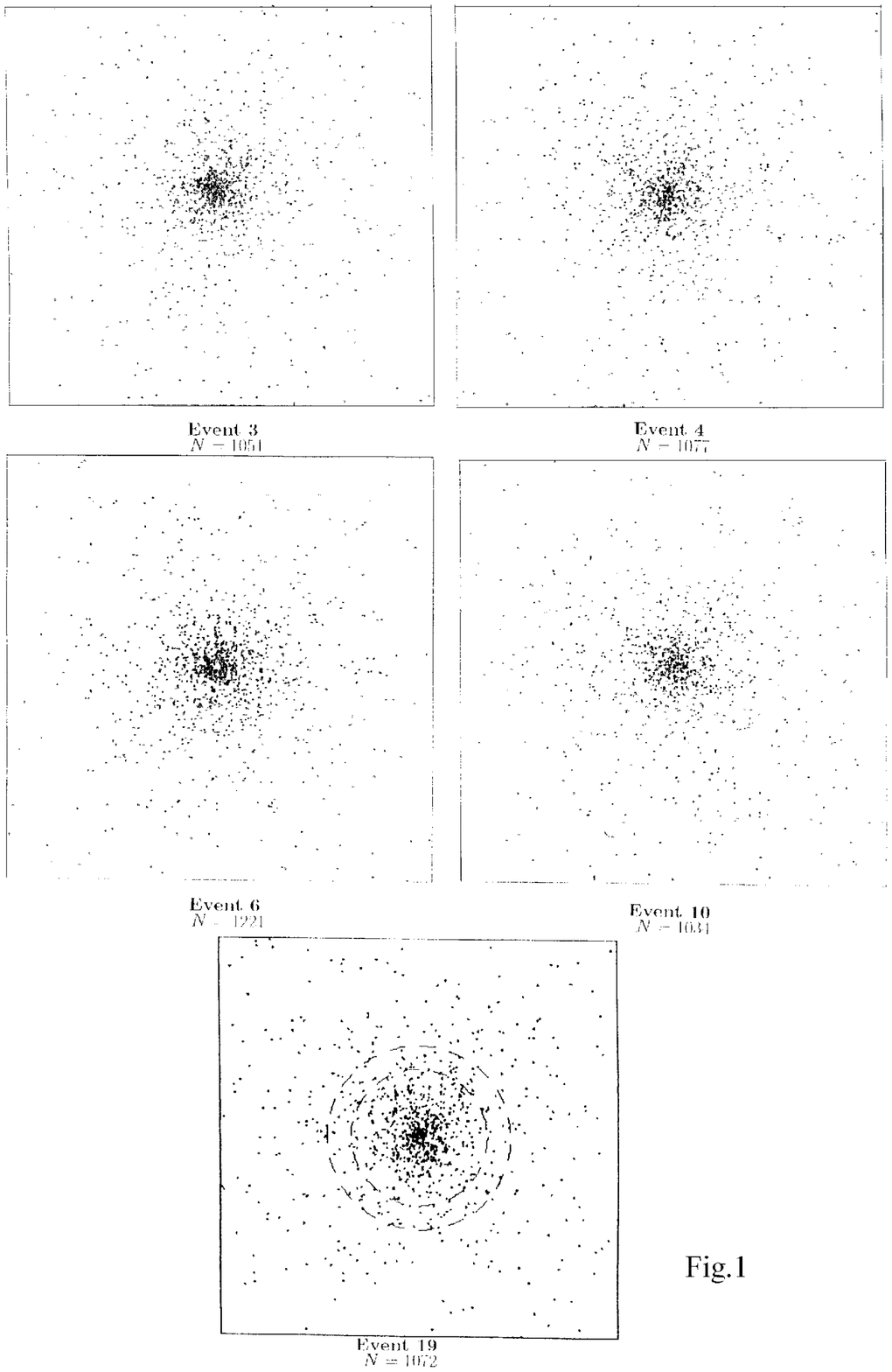,scale=0.65,clip=}
\end{center}

\vspace*{-1.5cm}
 The target diagrams of five events of central Pb-Pb collisions 
         at energy 158 GeV/nucleon obtained by EMU-15 collaboration.

\begin{center}
\vspace*{-2.5cm}
\epsfig{file=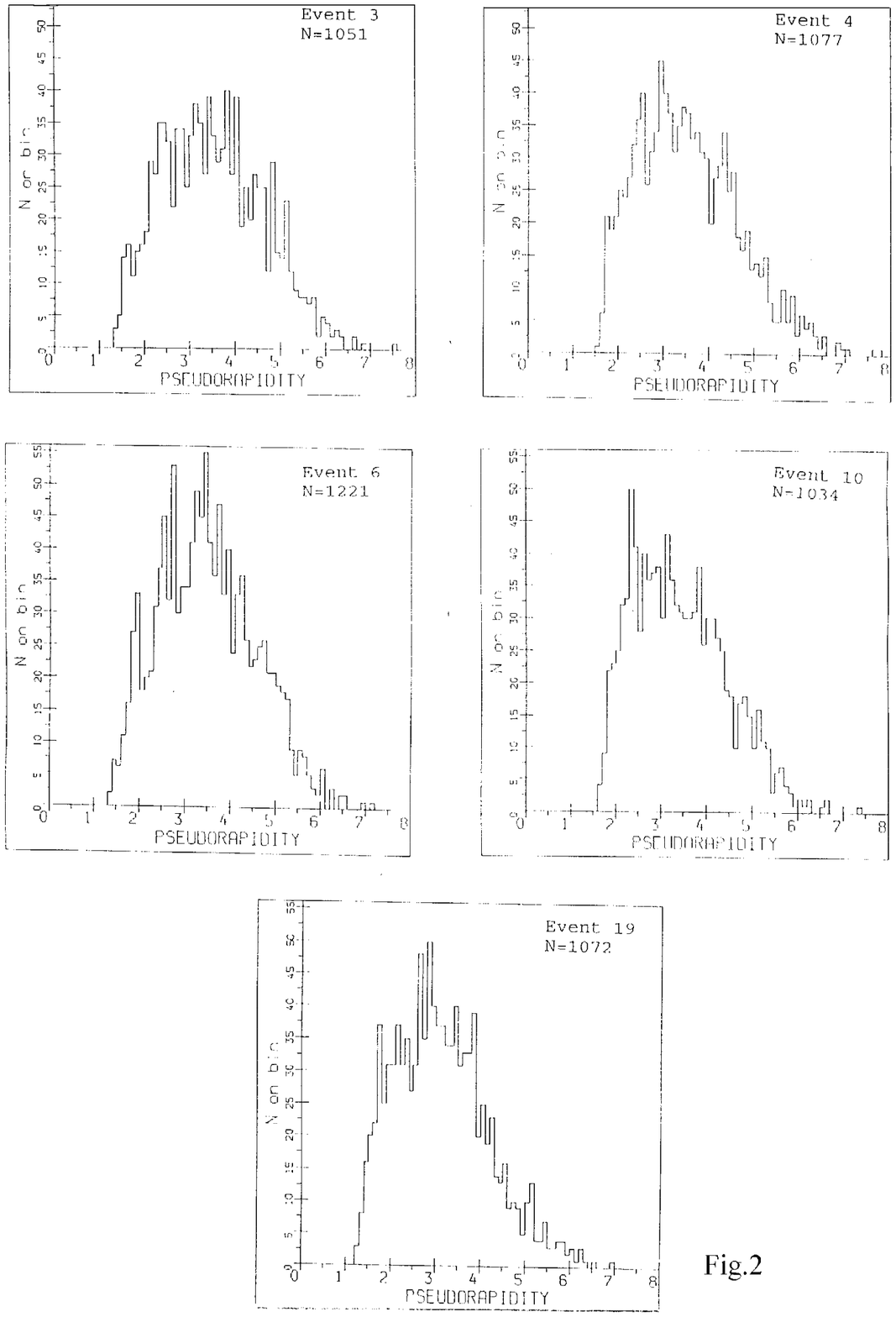,scale=0.7,clip=}
\end{center}
\vspace*{-1.5cm}
 The pseudorapidity distributions of particles in five events shown
         in Fig.1.\\

\begin{center}
\epsfig{file=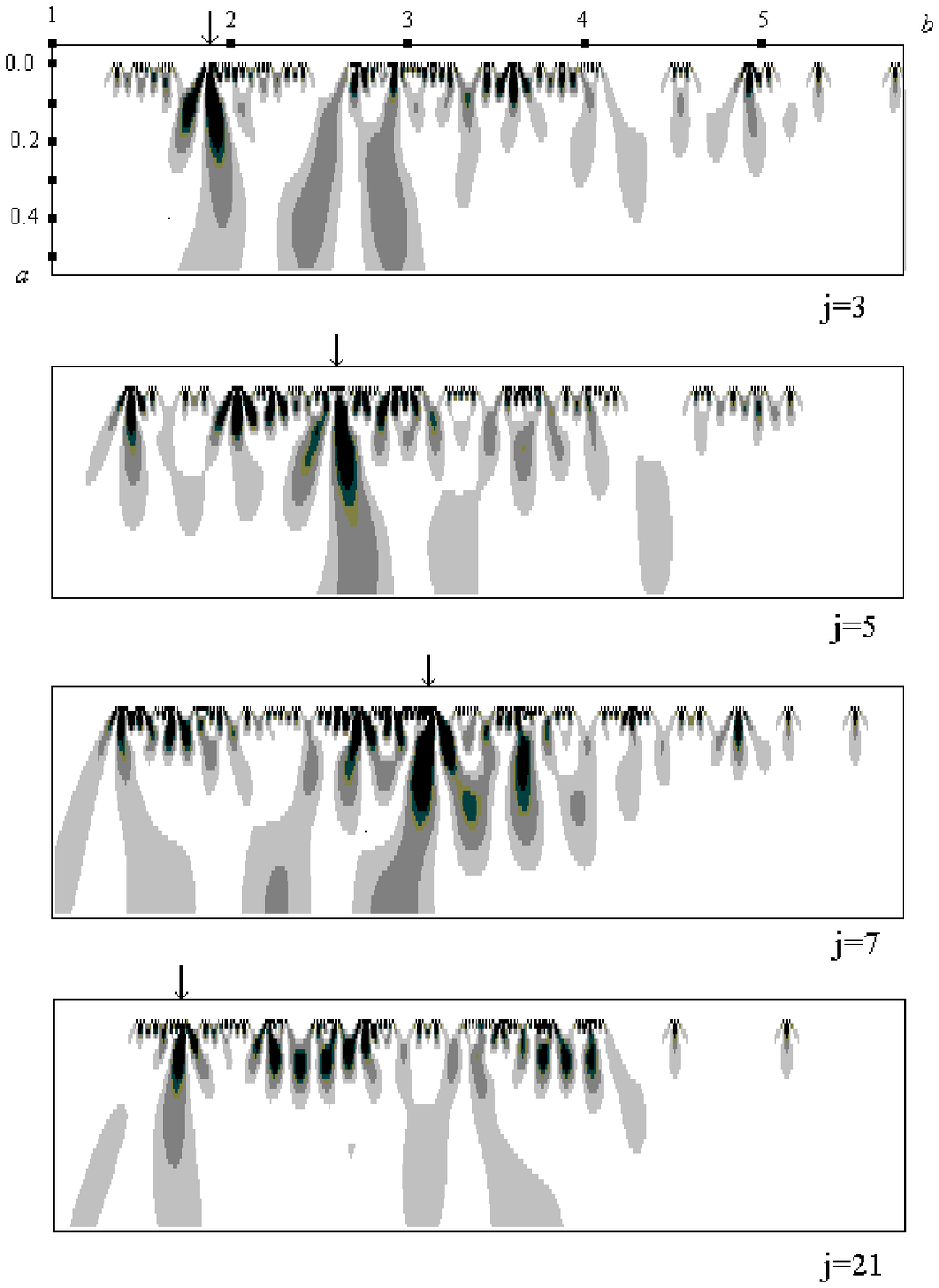, scale=0.6,clip=}%height=12cm}
\end{center}
Fig. 3   The wavelet coefficients for the event 19 analyzed in \cite{adk}.
        Dark regions correspond to large values of the coefficients.      
         Four of 24 sectors are shown. Rapidities are along x-axis,
         the scales increase down the vertical axis.

\begin{center}
\begin{tabular}{cc}
\fbox{\epsfig{file=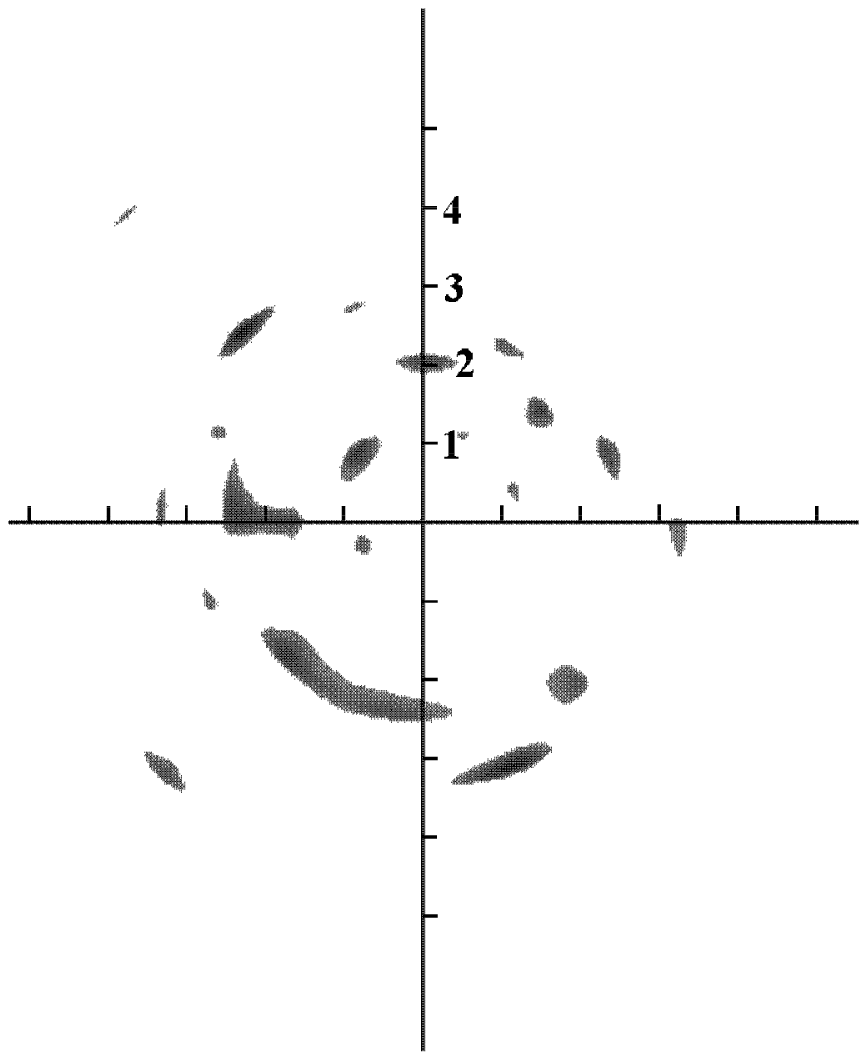,scale=0.47,clip=}}&
\fbox{\epsfig{file=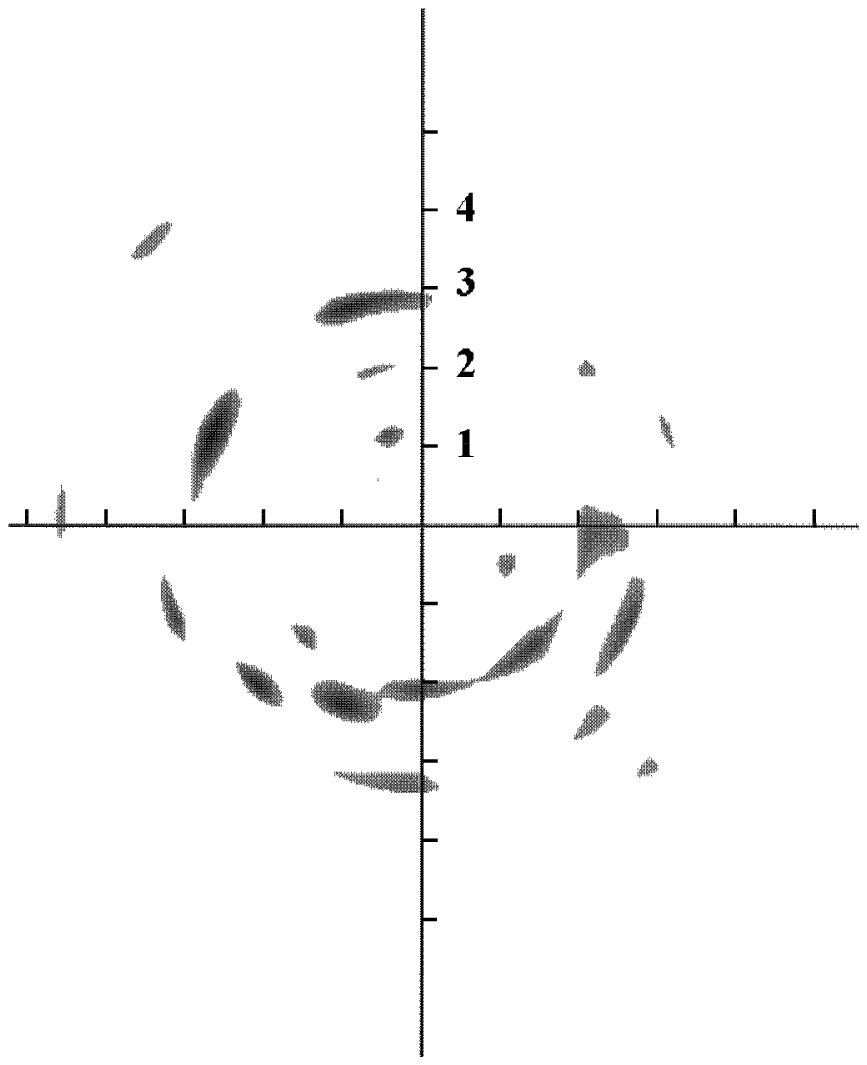,scale=0.47,clip=}}\
\end{tabular}
\end{center}
Fig.4 The modified large-scale target diagrams of two events (3 and 6).
      Darker regions correspond to larger particle density fluctuations. \\

\begin{center}
\fbox{\epsfig{file=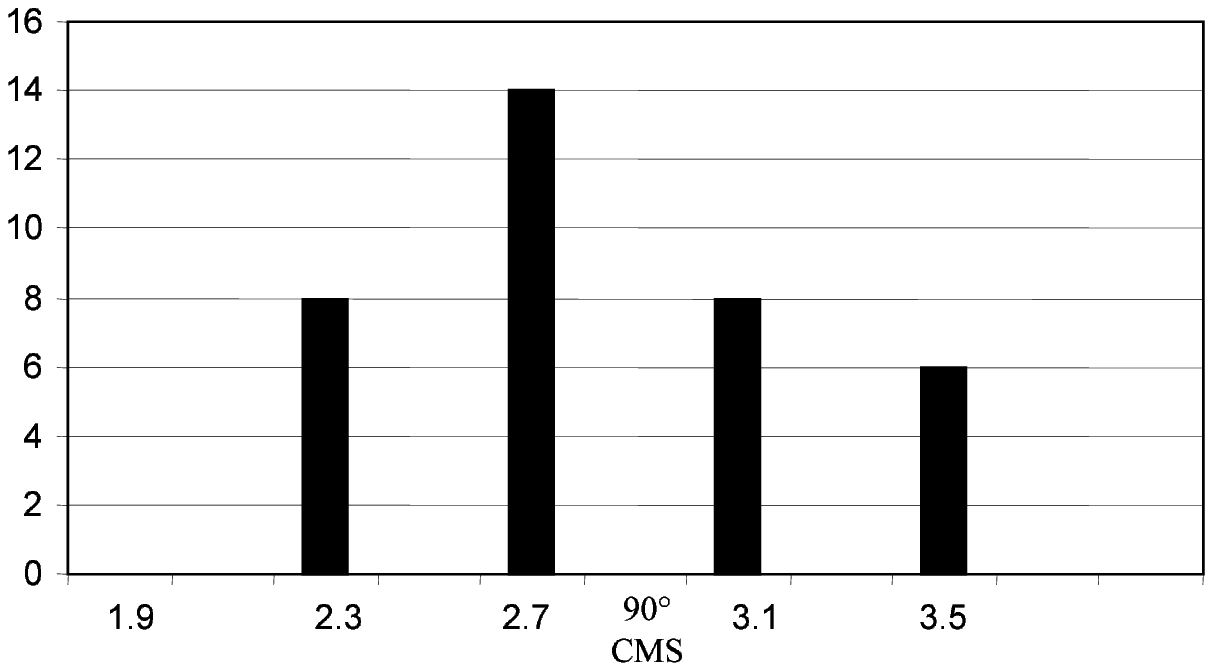,clip=}}
\end{center}
Fig.5 The pseudorapidity distribution of the maxima of \w\ coefficients.  \\
      The irregularity in the maxima positions, the empty voids between  \\
      them and absence of peaks at $\eta \approx 2.9$ are noticed. \\

\end{document}